\documentclass[a4paper,12pt]{article}

\usepackage [latin1]{inputenc}
\usepackage[brazil]{babel}
\usepackage[dvips]{graphicx}
\usepackage{geometry}
\usepackage{verbatim}
\usepackage{latexsym,amsfonts,amssymb}
\usepackage[normalem]{ulem}

\begin{document}
\noindent\textbf{\Large A speculative study of non-linear Arrhenius plot by using fractional calculus}\\

{\small\noindent{\bf  Nelson H. T. Lemes$^{a*}$, Valentino A. Simpao$^{b**}$, José P. C. dos Santos$^a$} \\

\noindent
\begin{flushleft}
 $^a$ 
Universidade Federal de Alfenas, Rua Gabriel Monteiro da Silva, 700 - Centro \linebreak  Alfenas/MG, 37130-000, Brazil\\

\noindent
$^b$ Mathematical Consultant Services, 108 Hopkinsville St. \linebreak Greenville, KY 42345, USA
\end{flushleft}


\vspace{\stretch{1}}

\noindent{$^*$nelson.lemes@unifal-mg.edu.br, $^{**}$mcs007@nuhlon.com}

\newpage
\subsubsection*{Abstract}
{In this study, the Van't Hoff differential equation is taken under consideration by making use of
fractional derivative tools. In this context, the nonlinear Arrhenius behaviour can be obtained and some
experimental values of reaction rate as function of temperature were fitted, with the  proposed  model. The new model showed
better performance to fit rate constant data for different kinetics process, when compared with
Arrhenius law. In these case, the Van't Hoff differential equation with noniteger order found relative percentage error
less that 3\% within experimental error. The fractional order plays an important role in modeling 
temperature dependence of these
kinetic processes. Thus it provides a new perspective in the handling of many problems 
(e.g., as solubility as function of temperature; temperature dependency of the viscosity and
conductivity, etc).}\\ 

\noindent
{\bf Keywords} non-linear Arrhenius plot; Van't Hoff equation; fractional derivative

\clearpage

\section{Introduction}

Experimental data show that reaction rates have a sensitive dependence on temperature; this point
lies at the beginning of chemical kinetics research.
In 1889 Svante Arrhenius \cite{Arrhenius}, 
by a thermodynamic argument due to
Van't Hoff \cite{Vanthoff}, 
proposed the 
simple equation to model 
effect of temperature on reaction rates. The empirical Arrhenius equation was applied
with great success in the modeling 
this influence by temperature on the rate of chemical reactions
(e.g., one of these reactions is the inversion of cane sugar by acids which
was discussed by Arrhenius in 1889). 
The Arrhenius equation describes that the rate of chemical reaction as decaying exponentially
with $1/T$ temperature; this relationship is known today as Arrhenius law.


Currently, with development of new
experimental techniques for studying
the influence of temperature
over a wide range and accurate experimental data, deviations from
Arrhenius law has been observed \cite{Masayoshi,Ratkowsky,Palmer}.
Therefore, for some reactions the temperature dependence of the rate constant is not exponential with $1/T$,
showing appreciable curvature of the Arrhenius semi-log plot of ($\ln k \times 1/T$);
these are called
non-linear Arrhenius behaviours.
This fault of the Arrhenius law can not be explained by experimental errors, 
wherefore new empirical equations has been used to model non-linear Arrhenius behaviour, such as
d-Arrhenius model which is inspired in Euler's approach to the exponential function \cite{Silva}.


The reference \cite{Laidler}
presents a review of the development of the Arrhenius equation, showing some of the extensions of Arrhenius law to cases
of deviation of Arrhenius plot from linearity.
Thus today it is  widely accepted that, over a limited range of temperature, the Arrhenius law for $\ln k$ as function as $1/T$
is acceptably linear but is not in general obeyed, for wide range of temperature.
There are a large number of theoretical models that have been proposed to interpret  this kind of process and
several
factors} considered
for deviations from linearity 
(e.g., quantum and collective effects \cite{Truhlar,Coutinho}).
As discussed in reference \cite{Coutinho}
the diffusion process plays an important role in effect of temperature
on reaction rate: this
motivated us to study the fractional calculus
in
this context.
Fractional calculus represents a natural instrument to model anomalous diffusion process,
when  memory effect have an important role therein \cite{Podlubny}.
One way to include this memory effect  is  to use fractional order derivatives in the original model, in this case
Van't Hoff equation.


Thus, we propose to analyze the
non-linear Arrhenius plot
in different way; by using
Van't Hoff differential generalized to non-integer order.
Recent studies 
have shown that
fractional calculus is a good alternative  tool
in many diverse fields of science,
such as
the study of the non-exponential growth of bacteria in culture media \cite{Zida}.
In reference \cite{Nelson}, the fractional differential equation
was used to model
an anomalous luminescence decay process for longer observation time, in which deviation
from exponential decay linear in the time variable is expected.
In  the  present  study,  for  the
first  time,  we  employ  Van't Hoff equation  modified with  the
fractional  derivative  to  obtain  the generalized model of Van't Hoff equation, from which
non-linear Arrhenius behavior can be explained. It will be shown that the curvature of
non-linear Arrhenius behaviour
can be
described by adjusting the fractional derivative order.


In the next section 2, we  will begin by introducing the necessary foundations of
fractional calculus and then our generalized model, the FVHE [Fractional Van't Hoff Equation]; it's exact analytical solution via operational methods is presented in section 3.
In section 4, numerical methods
which are also used to solve the FVHE are discussed.
Then in section 5,
experimental data {in a variety of contexts
will be compared}  with our proposed model (these include deformed Arrhenius law, curved Arrhenius plot and Arrhenius law),  showing that the FVHE
{is a good alternative model}
to describe these experimental data. 

\section{Fractional calculus background}


Theoretical description of fractional calculus begins by generalizing the integral with noninteger order, so if we
integrate $(Jf)(t)=\int_0^t f(s) ds$ {with respect to $s$ by}
$m$ times we find the following result \cite{Podlubny},
\begin{equation}
(JJJJ...JJf)(t)=(J^{m}f)(t)
=\frac{1}{(m-1)!}\int_0^t (t-s)^{m-1} f(s) ds
\label{eqint}
\end{equation}
Above equation is  well defined for all positive integers $m$. 

If we consider the fundamental theorem of calculus, 
in which $(DJf)(t)=f(t)$,  then we can write that
\begin{equation}
(D^mJ^mf)(t)=f(t)
\label{fund}
\end{equation}
by finite induction method, where $m$ is an integer.
In order to find a fractional
derivatives definition, take $m=n-\alpha$ and apply differential operator
$D^\alpha$ on both sides of equation (\ref{fund}) getting $(D^\alpha D^mJ^mf)(t)=(D^\alpha f)(t)$, thus it can be written as
\begin{equation}
(D^\alpha f)(t)=(D^nJ^{n-\alpha }f)(t)=\frac{1}{\Gamma(n-\alpha-1)}\frac{d^n}{dt^n}\int_0^t\frac{f(s)ds}{(t-s)^{\alpha -n+1}}
\label{eqdif}
\end{equation}
using composition rule \cite{Podlubny}.
Since equation (\ref{eqdif}) is well defined when $n-1\leq \alpha\leq n$ where
$n\in Z^+$ a positive integer, for $\alpha$ a noninteger number.

Thus one obtains a Riemann-Liouville fractional derivative, where $\alpha$  is defined as the fractional derivative order. This is in contrast to a derivative of integer order (a local operator), as the fractional operator is clearly non-local, since the fractional derivative depends on the lower boundary of the Riemann-Liouville integral. To make the notation clear we will use $D^\alpha_0$  as symbol of Riemann-Liouville fractional derivative.

\section{Fractional Van't Hoff differential equation}

Van't Hoff found the dependence of the equilibrium constant $K$ from the absolute temperature $T$, such as
$d(\ln (K(T)))/dT=\Delta H/RT^2$ \cite{Vanthoff}, where $\Delta H$ is the 
heat of reaction and $R$ is the gas constant.
From this equation and the relation of $K$ with the rate $k$, Arrhenius obtained
the expression  for the dependence of the rate $k$ with temperature absolute $T$ \cite{Arrhenius},
\begin{equation}
\frac{d(\ln (k(T)))}{dT}=\frac{E}{RT^2}
\label{vanthoffeq}
\end{equation}
in which 
$E$ is an empirically determined quantify called the activation energy.
Integrating above equation,
assuming $E$ is independent of temperature we get
\begin{equation}
\ln (k(T))=\ln A-\frac{E}{RT}
\label{arrh}
\end{equation}
where $\ln A$ is an integration constant.
Above equation is known as Arrhenius law in which
a plot of $\ln k$ versus $1/T$ will be linear with a negative slope equal to $-E/R$ and an intercept equal to $\ln A$.
As stated previously, this expression has had 
great success in describing the temperature dependence of reaction rates,
but for some reactions the behaviour of the plot of $\ln k \times 1/T$ is not linear.

In an attempt  to explain deviations from linearity  encountered in experimental results,
we propose a fractional-order generalization of the Van't Hoff equation: our FVHE, as
\begin{equation}
D_0^\alpha(\ln (k(T)))=\frac{E}{RT^2}
\label{neweq}
\end{equation}
in which $0\leq \alpha\leq 1$ is order of fractional Riemann-Liouville derivative. 
If the limit of the fractional order Limit $\alpha \rightarrow 1$, 
then 
the usual description of this process equation (\ref{vanthoffeq}) is recovered. 
{The curvature deviations of the  Arrhenius plot ($\ln k \times 1/T$),
according to our proposed model,
can be obtained 
by numerically solving above equation (\ref{neweq}) with Adams-Bashforth-Moulton  method \cite{Podlubny},
as discussed in section 4.}

For equation (\ref{eqdif}), we also attempt an operational solution:
Taking the Fourier transform of equation (\ref{neweq}) with respect to temperature $T$ mapping to
entropy $S$ yields
\begin{equation}
\begin{array}{c}
(-iS)^\alpha \breve{y}(S)=\frac{E}{R}\pi S\hspace{.2cm} {\rm sgn}(S)\\
\breve{y}(S)=\mathcal{F}_{T\rightarrow S}(\ln(k(T)))\\
\mathcal{F}_{T\rightarrow S}(\ln(k(T)))=\frac{1}{\sqrt{2\pi}}\int_{-\infty}^\infty\ln(k(T))e^{-iST}dT
\end{array}
\label{eq7}
\end{equation}
Although the particular value of the lower temperature limit in this work will be determined from experimental temperature data restrictions in the temperature domain for our subsequent analysis, said considerations do not restrict the domain of validity of equation (\ref{eq7})  being evaluated over the entire real line $\Re^1$, which of course includes  all real temperature domains. The Fourier transforms used were taken from \cite{Podlubny,Kammler}. Recall that temperature $T$  and entropy $S$   are conjugate thermodynamic variables; consequently the exact analytical solution to equation (\ref{neweq}) in entropy space may be expressed as
\begin{equation}
\breve{y}(S)=\frac{E}{R}\pi S (-iS)^{-\alpha}\hspace{.2cm} {\rm sgn}(S)
\label{yseq}
\end{equation}
where sgn(.) is sign function.
Hence via the inverse Fourier transform of (\ref{yseq}) from entropy space back to the temperature domain we obtain
\begin{equation}
\begin{array}{c}
\ln(k(T))=-\frac{E}{R}\mathcal{F}^{-1}_{S\rightarrow T}\{-\pi i(-iS)^{1-\alpha}\hspace{.2cm} {\rm sgn}(S)\}\\
\mathcal{F}^{-1}_{S\rightarrow T}(\breve{y}(S))=\frac{1}{\sqrt{2\pi}}\int_{-\infty}^\infty
\{-\pi i(-iS)^{1-\alpha}\hspace{.2cm} {\rm sgn}(S)\} e^{iST}dS
\end{array}
\label{eqk}
\end{equation}
Moreover, it follows that the operational exact analytical 
solution for temperature dependence rate coefficient $k(T)$ of the
FVHE
may thus be expressed as
\begin{equation}
k(T)=A\exp\left\{\frac{\Gamma(2 - \alpha) \cos(\pi \alpha)E}{RT^{2-\alpha}} \right\}
\label{final}
\end{equation}
in which $A=\ln k(T=T_0)$ is scaling constant with fractional order $\alpha \in (0,1)$ where $T>0$ with initial
temperature $T_0>0$ for temperature $T$ expressed in absolute Kelvin degrees.
Notwithstanding, this restriction  $T>0$ (T greater than absolute $0$ degrees Kelvin) presents no difficulty for our physical work, as there are no physical processes (e.g., chemical or otherwise) that occur at $T\leq 0$ degrees Kelvin: at $T=0$  (absolute $0$ Kelvin degrees): free bodies are still, no interaction within or without a thermodynamic system)
It follows that the result equation (10) is indeed valid for the entire physical work herein. We simply note that if for some reason a temperature scale other than Kelvin where chosen [e.g, Fahrenheit, Celsius or Rankine], then a different domain of the temperature variable $T \in \Re^1$  would be necessary, since in such cases, equation (\ref{final}) would then 'blow up' to infinity for $0$ degrees Fahrenheit, Celsius or Rankine; of course this is immediately remedied by simply re-scaling the temperature to absolute Kelvin units. Therefore the above result is obtained for the particular choice of $T_0>0$. As stated previously in the present Section 3, if the limit of Limit $\alpha\rightarrow 1$, then in such case equation (\ref{final}) is equivalent to equation (\ref{arrh}).


In order to discuss interpretation of the noninteger order $\alpha$ in our FVHE, as a consequence of equation
(\ref{eqk}) we have
\begin{equation}
\ln(k(T))=-\frac{E}{R}\mathcal{F}_{S\rightarrow T}^{-1}
\{\left( (-iS)^{1-\alpha}*\mathcal{F}_{T\rightarrow S}\{1/T\}\right)\}
\end{equation}
Now, considering that
$\mathcal{F}_{T\rightarrow S}\{J^\beta[f(T)]\}= \left((-iS)^{-\beta}*\mathcal{F}_{T\rightarrow S}\{f(T)\}\right)$,
as shown in reference \cite{Podlubny}, one obtains
\begin{equation}
\ln(k(T))=-\frac{E}{R}J^{\alpha-1}[1/T]
\label{eq12}
\end{equation}
in which $\beta=\alpha-1$ so that $J^{\alpha-1}[f(T)]$ is given by equation (1) with $m=\beta=\alpha-1$, $t$ replaced by $T$, and
$s$ still being a 'dummy variable' as in equation (1), which with these replacements becomes
\begin{equation}
(J^{\alpha-1})(T)=\frac{1}{\Gamma(\alpha)}\int_0^T\frac{f(s)}{(T-s)^{\alpha-2}}ds
\end{equation}
If $\alpha=1$  thus $J^{0}[1/T]=\frac{1}{T}$
as expected. 
Above equation (\ref{eq12}) is then
a 
convolution integral $\phi_{\alpha-1}*f$
in the range $[0,T]$ having memory kernel $\phi_{\alpha-1}=(T-S)^{\alpha-2}$ and
$f(T)\rightarrow f(s)=1/s$
in the above fractional integral, thus allowing an identification with retarded temperature dependence of the system (i.e., a non-local temperature phenomenon). Notwithstanding the zero lower bound in the fractional integral above, the solution
 $k(T)$ of equation (10) for our physical work is only being considered for  the temperature domain interval from initial non-zero temperature  $T_0$ to infinity (i.e.,[$T_0$,$\infty$), $\ni$ $T_0=0$).

 In the present report, this concludes our introductory sketch of the analytical solution to the FVHE. In the remainder of this work, we shall primarily utilise numerical solution techniques for  the FVHE, to which we now turn.

\section{Numerical approach for fractional differential equation}

In this section, we describe the numerical method based on trapezoidal rule to solve fractional integral equation (1).  Suppose that interval $[0,t]$ is subdivided into $n$ between $s_k=kh$ and $s_{k+1}=(k+1)h$ in which $h=t/n$,
\begin{equation}
J^\alpha f(t) =\frac{1}{\Gamma(\alpha)}\sum_{k=0}^{n-1}\left[\int_{kh}^{(k+1)h}(t-s)^{\alpha-1}f(s)ds\right]
\end{equation}
\noindent
The  function $f(s)$ between $s_k$ and $s_{k+1}$ is given by linear interpolation,
such as, $f(s)\approx f(kh)+\frac{\Delta f}{h}(s-kh)$ in which
$\Delta f=f((k+1)h)-f(kh)$. Therefore,  equation (13) can be rewritten as the following equation
\begin{equation}
J^\alpha f(t) =
\frac{1}{\Gamma(\alpha)}\sum_{k=0}^{n-1}\left[f(kh)w_{n-k}^\alpha
+\frac{\Delta f}{h}g_{n-k}^\alpha\right]
\end{equation}
in which  
\begin{equation}
w_{n-k}^\alpha =\int_{kh}^{(k+1)h}(t-s)^{\alpha-1}ds=\frac{h^\alpha}{\alpha}   [i^\alpha-(i-1)^\alpha]
\end{equation}
with $i=n-k$, and
\begin{equation}
g_{n-k}^\alpha =\int_{kh}^{(k+1)h}(s-kh)(t-s)^{\alpha-1}ds=
\frac{h^{\alpha+1}}{\alpha(\alpha+1)}[
i^{\alpha+1}-(i+\alpha)(i-1)^\alpha]
\end{equation}
The details of this method can be found in reference \cite{Diethelm}. This way can be generalized to interval $[a,t^\prime]$ and result used to solve a fractional differential equation, as for our VFHE equation (6) where $a\rightarrow T_0$, $t\rightarrow T$ for $T>0$ and $T_0>0$.

The
FVHE (6) can be generalized as
$D_a^\alpha y(t)=f(t)$
with initial conditions $y(a)=y_a$ which is necessary when $0<\alpha<1$. The initial value problem (16) is equivalent
to following equation
\begin{equation}
y(t)=y_a+J_{a}^\alpha f
\end{equation}
which is a generalization of Adams-Bashforth-Moulton method to fractional differential equation \cite{Podlubny}.
Therefore knowing  $J_{a}^\alpha f$ from equation (14) we can find $y(t)$ for all time. The symbol $J_{a}^\alpha f$ represents
fractional integral equation on interval between $a$ and $t$.




The $\alpha$ parameter is estimated by fitting the result of equation (\ref{neweq}) to the experimental data using
the graphical
method
that minimizes the sum of squares of the residual error,
$E=1/2\sum_i(\ln k_i^{cal}-\ln k_i^{exp})^2$. {The
aim of this paper does not involve only a
curve fitting problem,
but the success of our FVHE shows that fractional calculus may provides
a new insight to these phenomena.}

\section{Results and discussions}

In this section we show that the usual assumption of Arrhenius plot is not valid in many cases.
The first example reports study of thermal decomposition of Diacetylene between 973 e 1223 K \cite{Palmer}.
The next example
deals with experimental data of the respiration rate for $O_2$ consumption by leaves of
{\it Camellia Japonica} \cite{Masayoshi}. Finally, some cases to bacterial growth will be studied \cite{Ratkowsky}.

When there are significant inconsistencies between 
experimentally measured
and theoretically calculated rate constant, calculated values can be brought into agreement with  experimental values
by making use of the fractional
differential order. As shows in figure 1, the $\alpha$ parameter play an important role in temperature
dependence of rate constant. When $\alpha$ decreases the rate constant increases with temperature, but
this increase is less than observed for the $\alpha=1$.
In the following,
the proposed model will be compared with Arrhenius, d-Arrhenius, quadratic equations and experimental data cited.

The best parameter $\alpha$  was found minimizing the sum of squared residuals by using graphical technique. The Figure 2 shows the
relative percentage error as function as fractional order $\alpha$, obtained to thermal decomposition kinetics of
$C_4H_4$ \cite{Palmer}. The minimum value in this curve was used as the best value of $\alpha$, 0.734.
The results from best fractional order when used in equation (6) to
fit experimental data
are shown in Table 1. {Table 2 presents the parameters $E/R$ to the best fractional order together with $E/R$ obtained from adjust of Arrhenius equation. Table 2 too includes the experimental initial conditions $y_a$.}

The code developed by  Garrappa \cite{Garrappa}
was used in this paper to solve an initial value problem for a nonlinear fractional differential equation (6). This implementation is based in predictor-corrector method  of Adams-Bashforth-Moulton, as described in same reference.


As shown in Figure 3
experimental data retired from reference \cite{Palmer} presents a concave curve in Arrhenius plot, with curvature
$\kappa=-3.7\times 10^7 $.
The similar behavior is seen in all experimental data studied here \cite{Palmer,Masayoshi,Ratkowsky}.
The performance of the fractional Van't Hoff equation
to fit experimental data
was assessed according to the relative percentage error criteria, as shown in Table 1.
Results from equation (6) were compared with
three different models: Arrhenius, d-Arrhenius and Quadratic polynomial.

The fractional Van't Hoff equation to provide a
good alternative to fit the experimental temperature dependence data, when compared with usual  Van't Hoff equation
was obtained
a relative percentage mean error of
1.9 \% against 6.7 \%.
This success is due to the flexibility included by
fractional order in Van't Hoff differential equation. In some cases, equation (6) presents performance better than
quadratic equation.  When compared with d-Arrhenius equation, the equation (6) shows similar relative percentage errors,  with relative percentage mean error of 1.9 \% against 1.5\%. {Therefore, it is believed that our 
generalized model of the Van't Hoff equation
{provides a new perspective on interpretation} of  
temperature dependence of the rate constant
at the microscopic level.}





\section{Conclusions}

In this paper, was shown that the Arrhenius law fails in describing  the temperature dependence of rate coefficients for
some kinetics process, with
relative percentage error
greater than 10 \%, which is greater than experimental error. In these cases, the generalized Van't Hoff differential equation herein, can be used with good results to fit experimental kinetics data within experimental error
with relative percentage error
less that 3 \%. The proposed model finds
results consistent with the d-Arrhenius model by using of different tools [i.e., fractional derivatives]. The d-Arrhenius model uses Euler approach to exponential function in Arrhenius law, with additional parameter $d$. While our FVHE model includes a fractional order $\alpha$  instead of $d$ in the most fundamental equation from which Arrhenius law is derived,
there has been no correlation  observed  between  $\alpha$ and $d$ parameters; as well as any correlation between $\kappa$ and $\alpha$ parameters, for examples discussed in this paper.
This approach can be applied to fit various other experimental data such as: solubility as function of temperature; temperature dependency of the viscosity and
conductivity, etc.

\section*{Acknowledgments}

We would like to thank 
FAPEMIG for their financial support.

\clearpage
\begin{center}
\begin{figure}[h]
\centering
\includegraphics[scale=.4]{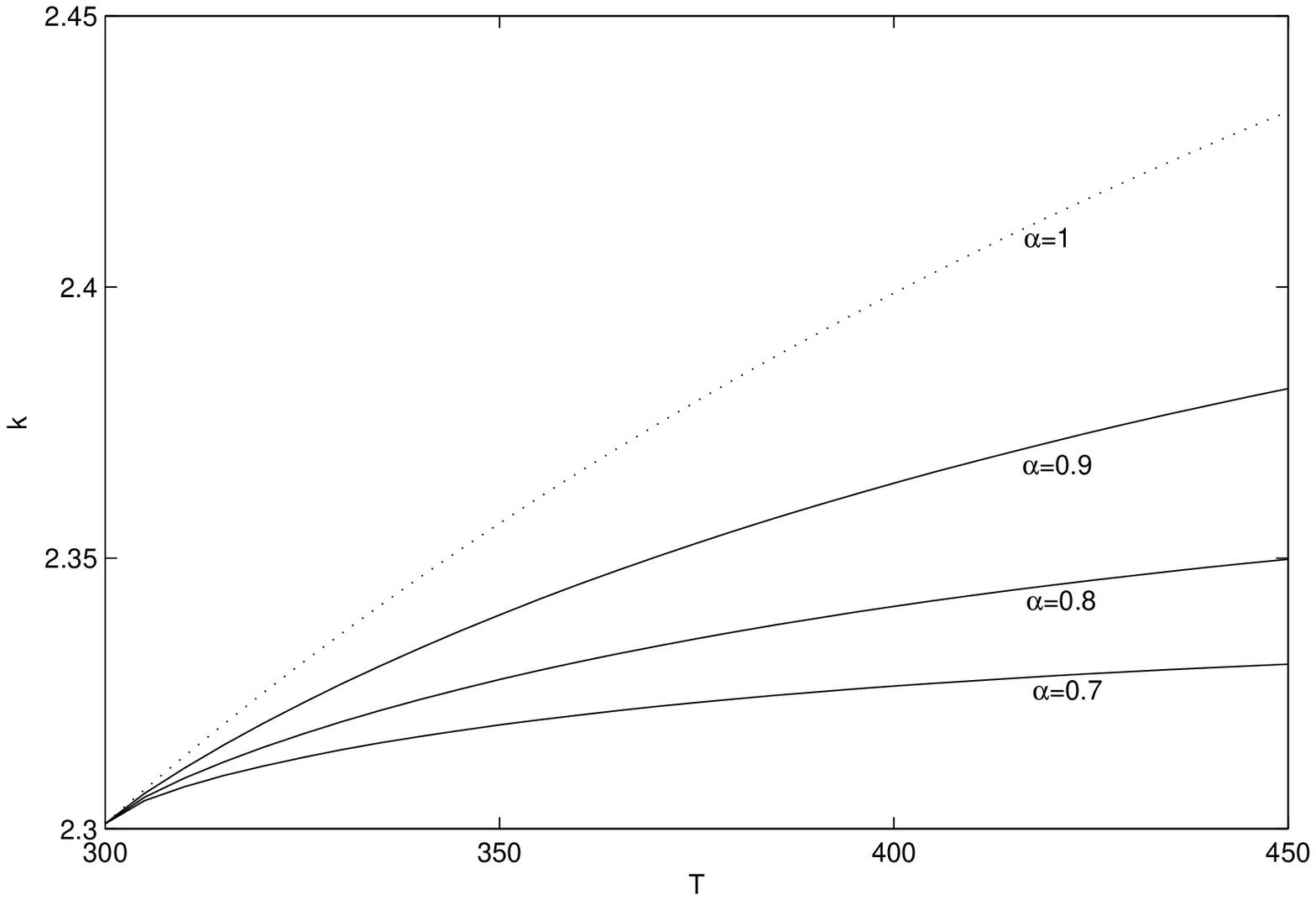}
\includegraphics[scale=.4]{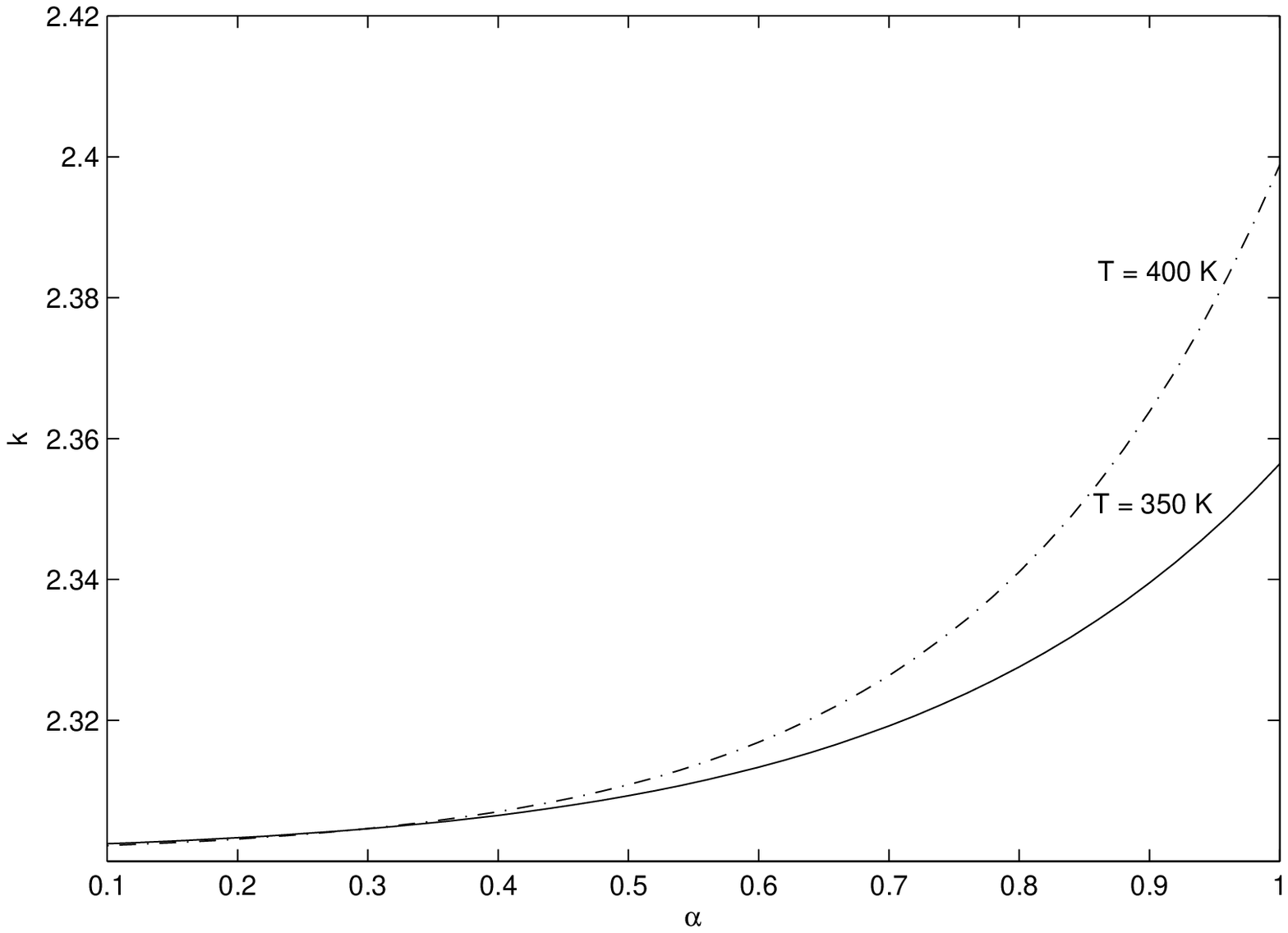}
\caption{
Simulated data using equation (6) with E/R=50 arbitrary unit.
}
\label{figure1}
\end{figure}
\end{center}

\clearpage
\begin{table}[!h]
\caption[curto2]{Comparison between  Arrhenius model, this work, d-Arrhenius model and quadratic function.}
\label{teste3}
\begin{center}
\begin{tabular}{c|ccc|cccc}
\hline\hline
exp. data & $\kappa^{(f)}$ & $\alpha_{\rm best}$ & $d_{\rm best}$ & relative error, \% & & & \\
          &      &          &     & Arrhenius & This work & d-Arrhenius$^{(d)}$ & Quadratic$^{(e)}$ \\
\hline
\cite{Palmer}      & -3.7$\times 10^7$     & 0.734 &  0.389 &   0.9739  & 0.2737 &  0.2530 &  0.3309\\
\cite{Masayoshi}      &  -1.4$\times 10^7$    & 0.750 &  0.228 &   1.8636  & 1.0663 &  0.8938 &  0.6077\\
\cite{Ratkowsky}$^{(a)}$      & -3.9$\times 10^7$     & 0.698 &  0.142 &   11.7670  & 4.0697 &  3.0871 &  2.1995\\
\cite{Ratkowsky}$^{(b)}$      &  -7.5$\times 10^7$    & 0.830 &  0.200 &   3.5652 & 2.6406  & 2.4432  & 2.0792 \\
\cite{Ratkowsky}$^{(c)}$      &  -1.0$\times 10^8$    & 0.572 & 0.236 & 15.3584 & 1.4083 & 0.8900 & 4.4934\\
\hline
\hline
\end{tabular}\end{center}\vspace{0.1cm}
{\small ${(a)}$ Aerobacter aerogenes, ${(b)}$ Bacillus circulans, ${(c)}$ Lactobacillus delbruckil}\\
{\small ${(d)}$ $\ln k=\ln A +\frac{1}{d}\ln\left(1-d\frac{E}{RT}\right)$},
{\small ${(e)}$ $\ln k=\ln A +B\frac{1}{T}+C\left(\frac{1}{T}\right)^2$},
{\small ${(f)}$ $\kappa=2C$}\\
\end{table}

\clearpage
\begin{table}[!h]
\caption[curto3]{Initial conditions of equation (6) and activation energy empirically determined .\\}
\label{teste4}
\begin{center}
\begin{tabular}{c|ccc|cc}
\hline\hline
          & eq. (6)     &          &   & eq. (5) &   \\
exp. data & $E/R$ & $y_a$ & $a$,K & $E/R$ & $\ln A$ \\
\hline
\cite{Palmer}      & 5.87$\times 10^4$     & 16.15     &  973 &  6.01$\times 10^3$   & 13.36 \\
\cite{Masayoshi}      &  2.41$\times 10^4$    & -15.09 &  270 &   4.54$\times 10^3$  & 10.41 \\
\cite{Ratkowsky}$^{(a)}$      & 5.92$\times 10^4$  & 2.39    & 274 &  9.34$\times 10^3$ &     35.52\\
\cite{Ratkowsky}$^{(b)}$      &  4.00$\times 10^4$ & 6.46   & 311 &  1.11$\times 10^4$ &     38.66 \\
\cite{Ratkowsky}$^{(c)}$      &  1.23$\times 10^5$ & 2.11   & 299 & 1.36$\times 10^4$ &   46.96\\
\hline
\hline
\end{tabular}
\end{center}
\end{table}

\clearpage
\begin{center}
\begin{figure}[h]
\centering
\includegraphics[scale=.8]{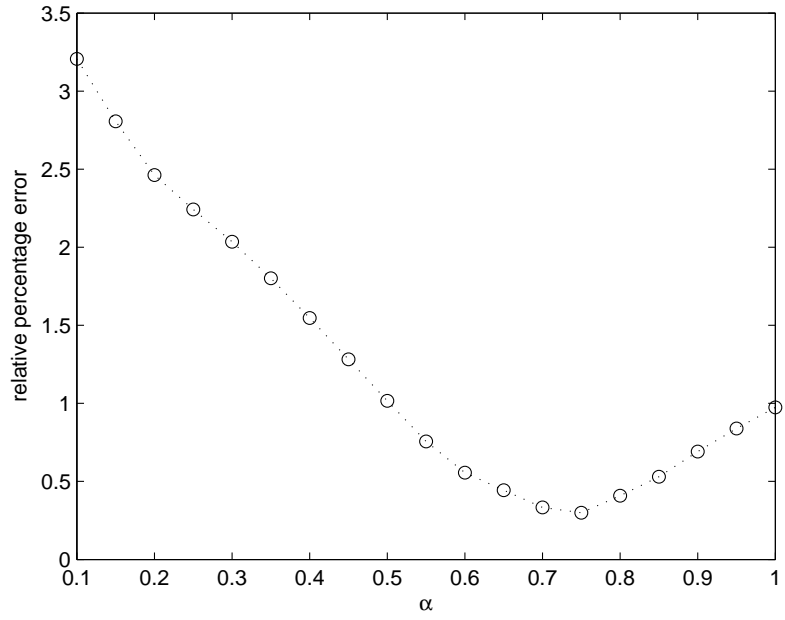}
\caption{
Relative percentage error as function of fractional differential order.
}
\label{figure1}
\end{figure}
\end{center}

\clearpage
\begin{center}
\begin{figure}[h]
\centering
\includegraphics[scale=.8]{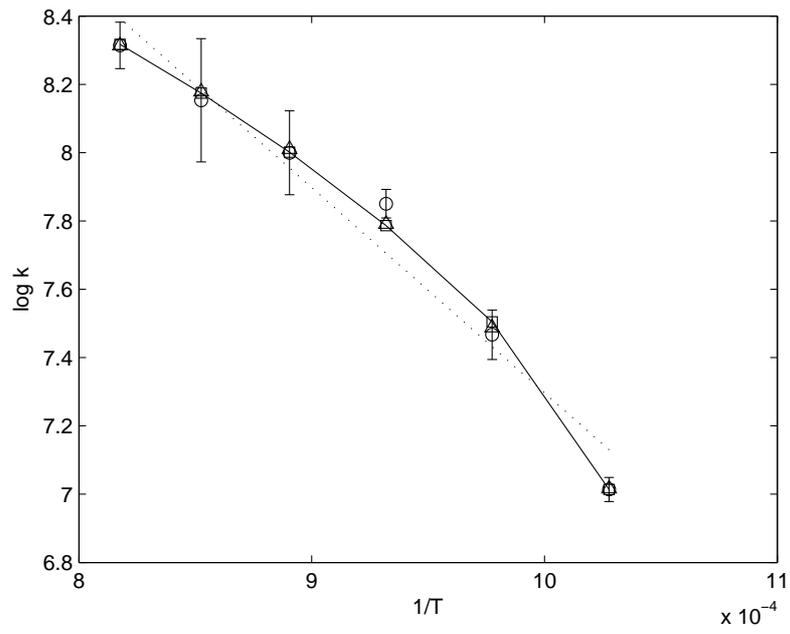}
\caption{Comparison between  Arrhenius model (dotted line), this work (square and continued line), d-Arrhenius model (triangle) and experimental data (circle).}
\label{figure1}
\end{figure}
\end{center}

\end{document}